\documentclass[11pt]{article}
\usepackage[margin=1in]{geometry}
\usepackage{times}
\usepackage[T1]{fontenc}
\usepackage{microtype}
\usepackage{booktabs}
\usepackage{amsmath}
\usepackage{graphicx}
\usepackage{pgfplots}
\pgfplotsset{compat=1.16}
\usepackage[numbers,sort&compress]{natbib}
\usepackage[hidelinks]{hyperref}
\usepackage{xcolor}

\makeatletter
\renewcommand\paragraph{\@startsection{paragraph}{4}{\z@}{1.0ex \@plus .3ex \@minus .2ex}{-1em}{\normalfont\normalsize\bfseries}}
\makeatother
\newcommand{\eb}{Earlybird}
\newcommand{\ms}{\,ms}

\title{\eb{}: Predictive Streaming Speech Recognition in 29M Parameters}
\author{Sahil Mahendrakar}
\date{November 2026}

\begin{document}
\maketitle

\begin{abstract}
Streaming speech recognizers output a word only after the speaker has finished it, typically
90--400\ms{} later. For a voice agent every millisecond counts: the sooner the words are outputted, the
sooner the agent can start a search or plan its reply. We present \eb{}, a 29M-parameter streaming
recognizer (16.8\,MB at 4 bits). \eb{} learns from audio cut off at
random points, including partway through a word, and is asked to predict every word that is due by
then in full, plus a guess of the next two words. Its raw token probabilities give a confidence for
every word including the one still being spoken. On clean audiobook speech (LibriSpeech
test-clean), the median word is predicted correctly 90\ms{}
before the speaker finishes it, while released streaming models get it 90 to 400\ms{} after it
ends. \eb{} is the earliest on all six test sets we time, covering audiobooks, TED talks, public
speech, earnings calls, and synthetic speech. On the same audiobooks, 69\% of words are predicted
correctly before they end, against 3--25\% for the other models. For words still being spoken,
confidence is calibrated, with an expected calibration error (ECE) of 3.7 to 5.5. On the
eight English Open ASR Leaderboard sets, it is more accurate than Moonshine Tiny on every set and
roughly ties Whistle, a model of the same size.
\end{abstract}

\section{Introduction}

A spoken request to a voice agent reaches the agent as text. With a streaming recognizer, each
word of that text arrives a few hundred milliseconds after the speaker finished saying it.
For a voice agent, that delay is time it could have spent starting a search or
planning a reply while the speaker is still talking.

People listening to speech do not wait for the end of a word to know what it is. Most words can
be recognized from their first few phonemes in context, and the next word is often predictable.
We ask whether a small on-device recognizer can do the same: predict each word from its first
sounds, guess the word after it, and say how sure it is.

\eb{} trains on utterances cut off at random 80\ms{} steps, which is exactly the position a
streaming model is in at each update. The target is every word that is \emph{due} at that point,
written in full even if its audio has only begun, followed by an end marker and the next two
words of the transcript as a guess. Because the targets for successive cut points only ever extend
one another, the model also learns not to output what it will later take back. Every word it outputs
carries the product of its token probabilities as a confidence.

Our contributions are:
\begin{itemize}
  \item A training recipe that turns a released streaming recognizer (Moonshine streaming
        Tiny~\citep{jeffries2024moonshine,moonshinestreaming}) into one that predicts words correctly before
        they end, with no change to the architecture and no future audio
        (Section~\ref{sec:method}).
  \item A measurement of earliness, stability and accuracy against released streaming models run
        with their own code, on six test sets with per-word timings (Section~\ref{sec:early}).
  \item A 16.8\,MB 4-bit release scored on the Open ASR Leaderboard's English sets
        (Section~\ref{sec:openasr}).
  \item Evidence that the model's raw confidence is calibrated for words still being spoken and
        for guessed next words (Section~\ref{sec:conf}).
\end{itemize}

\section{Related work}

\paragraph{Predicting words before they are spoken.}
\citet{zink2025predictive} train an offline encoder-decoder to predict words in masked audio at
the end of an utterance (``predictive speech recognition'') and report useful predictions up to
about 300\ms{} before the utterance ends. \citet{yusuf2024speculative} define speculative speech
recognition, in which a language model prompted with audio runs ahead of a streaming recognizer.
\citet{baumann2011microtiming} predict the remaining duration of the current word for an
incremental dialogue system. \eb{} differs in putting prediction inside a small streaming recognizer,
training on random cut points rather than only the utterance end, and reporting calibration and
stability alongside earliness.

\paragraph{Using early partial results.}
\citet{chang2020prefetch} prefetch downstream responses from partial hypotheses and save about
200\ms{} when the final result matches. \citet{schwarz2023personalized} predict whole utterances
from partial ones to cut assistant latency. Both depend on how early and how stable partial text
is, which \eb{} improves.

\paragraph{Low-latency streaming recognition.}
Work such as FastEmit~\citep{yu2021fastemit} reduces how long a transducer waits before emitting a
token, and LocalAgreement~\citep{machacek2023whisperstreaming} turns an offline model into a streaming
one by committing text that consecutive passes agree on. These methods reduce delay toward the end
of the word. \eb{} aims to predict the word correctly before it ends.

\section{Method}
\label{sec:method}

\subsection{Starting point}
\eb{} starts from Moonshine streaming Tiny~\citep{moonshinestreaming} (34M parameters, MIT
license), an encoder-decoder that consumes 80\ms{} audio chunks. We cut its vocabulary to 19,007
tokens: those that occur in a large body of English text (our training transcripts), plus every
single-character token, so any word can still be spelled out. This brings it to 29M parameters,
and the cut model scores within 0.4 WER points of the original on every test set, including
TED-LIUM and Earnings-22, which are not in its training data. The released runtime holds each word back until roughly 300\ms{} of audio past it has
been heard. \eb{} decodes the full transcript at every step with no hold-back.

\subsection{Training on cut-off audio}
Let an utterance have words $w_1,\dots,w_n$ with start and end times $s_i$ and $e_i$ from a forced
aligner. Word $w_i$ is due at
\begin{equation}
  \tau_i = \max\bigl(s_i + \min(0.5\,(e_i - s_i),\, h),\; e_i - \ell\bigr),
\end{equation}
with lead $\ell = 300$\ms{} and minimum heard audio $h = 50$\ms{}: a word of ordinary length is due
300\ms{} before it ends, and a short word once some of it has been heard. Each training example is
the utterance cut at a random 80\ms{} step $t$. The target is every word with $\tau_i \le t$, spelled as
in the transcript, then an end-of-text token, then the next two words as a guess. Trailing
punctuation is held back until the next word is due, so a comma never appears and disappears.
Since the target at $t' > t$ always extends the target at $t$, the model is never trained to write
something it will later remove.

The guessed words follow the end-of-text token, so a decoder that stops at end-of-text produces
only the transcript. Continuing past it produces the guess.

\subsection{Learning from the teacher only where it is right}
The original Moonshine Tiny also serves as a teacher through a distillation loss, applied only at
token positions where the teacher's top choice matches the label. Where the teacher is wrong or
unsure, its probability is spread over several words, and matching it there taught \eb{} to do the
same. With the loss at every position, the guessed words came out less decisive, with their
probability spread over several candidates instead of committed to one.

\subsection{Padding frames}
At each step the encoder also sees three frames computed from silence after the audio heard so
far. This uses no future audio. It lets the model finish a word as if the speaker might have
stopped. Training with padding frames moved the median word from $-50$\ms{} to $-90$\ms{} on
LibriSpeech test-clean (Table~\ref{tab:ablate}).

\subsection{Quantization}
Quantization-aware training brings the weights to 4 bits, with the audio front end kept at 8 bits.
The weights file is 16.8\,MB (17.6\,MB with the trimmed tokenizer and configuration).

\subsection{Confidence}
For each word on screen, the confidence is the product of the probabilities of its tokens in the
decoder pass that produced it. Guessed words get their confidence the same way. There is no
temperature scaling or other calibration step.

\subsection{Data and schedule}
Training uses LibriSpeech~\citep{panayotov2015librispeech} (960\,h, published Montreal Forced
Aligner~\citep{mcauliffe2017mfa} word times), about 300\,h of People's Speech
clean~\citep{galvez2021peoples} with word times from a wav2vec~2.0 CTC aligner, and about 25\,h of
synthetic Kokoro~\citep{kokoro} speech whose phoneme durations give exact word times, sampled with
weights 1:3:3:3:1 (LibriSpeech clean-100, clean-360, other-500, People's Speech, Kokoro; the first
stage below uses the four real sets only, 1:3:3:3). Real
transcripts are kept only when a teacher (Moonshine Medium) writes exactly the reference words.
No non-commercial corpora are used for training. Training ran in four stages on a single GPU with
batch 48, all with the KD term: 10k steps of streaming-cut training after the vocabulary cut (learning
rate $5\times10^{-5}$), 20k steps of prefix training with two guessed words ($5\times10^{-5}$), 5k
steps with the padding frames ($2\times10^{-5}$), and 5k steps of quantization-aware training at
4 bits with the front end at 8 bits ($2\times10^{-5}$).

\section{Evaluation setup}

\paragraph{Test sets.} LibriSpeech test-clean and test-other, TED-LIUM 3~\citep{hernandez2018tedlium},
a held-out People's Speech set, Earnings-22~\citep{delrio2022earnings}, and 500 held-out WikiText
sentences synthesized with Kokoro using two voices not seen in training. Word end times come from
the Montreal Forced Aligner (LibriSpeech), Kokoro's durations (synthetic), or our wav2vec~2.0
aligner, whose word ends fall within a median of 4--8\ms{} of Kokoro's and 26\ms{} of MFA's.

\paragraph{Harness.} Each recognizer is fed audio in order in 80\ms{} chunks and every change to
its on-screen text is logged against the audio clock. Processing time is not added for any model,
so the numbers measure what each model decides to output, not how fast the hardware runs it. For
each reference word we record the time from which it is \emph{predicted correctly}: the word is in the
output, spelled correctly and in the right location, and stays unchanged until the end of the utterance. Its \emph{word latency} is that time relative to the word's acoustic end; negative means before the
speaker finishes it. We also count revisions (words taken back) per spoken word and word error rate
(WER) of the final transcript. Confidence is scored with expected calibration error
(ECE)~\citep{guo2017calibration} over 10 equal-width bins. Every model is scored on the same utterances within each set. Unless stated otherwise, \eb{}
results are for the full-precision model; the 4-bit release is in Tables~\ref{tab:openasr}
and~\ref{tab:ablate}.

\paragraph{Baselines.} Released models run with their own code: Moonshine Tiny with its official
streaming runtime (quantized); Whistle~\citep{whistle} (55M parameters, 16.9\,MB) in the Cactus
engine's live mode; and Whisper base.en~\citep{radford2023whisper} (74M) re-decoded on all audio so
far every 0.48\,s, with its raw output shown. For accuracy and on LibriSpeech clean we also run
Whisper tiny.en (39M) and small.en (244M) the same way.

\section{Results}

\subsection{Earliness, accuracy and size}
\label{sec:pareto}

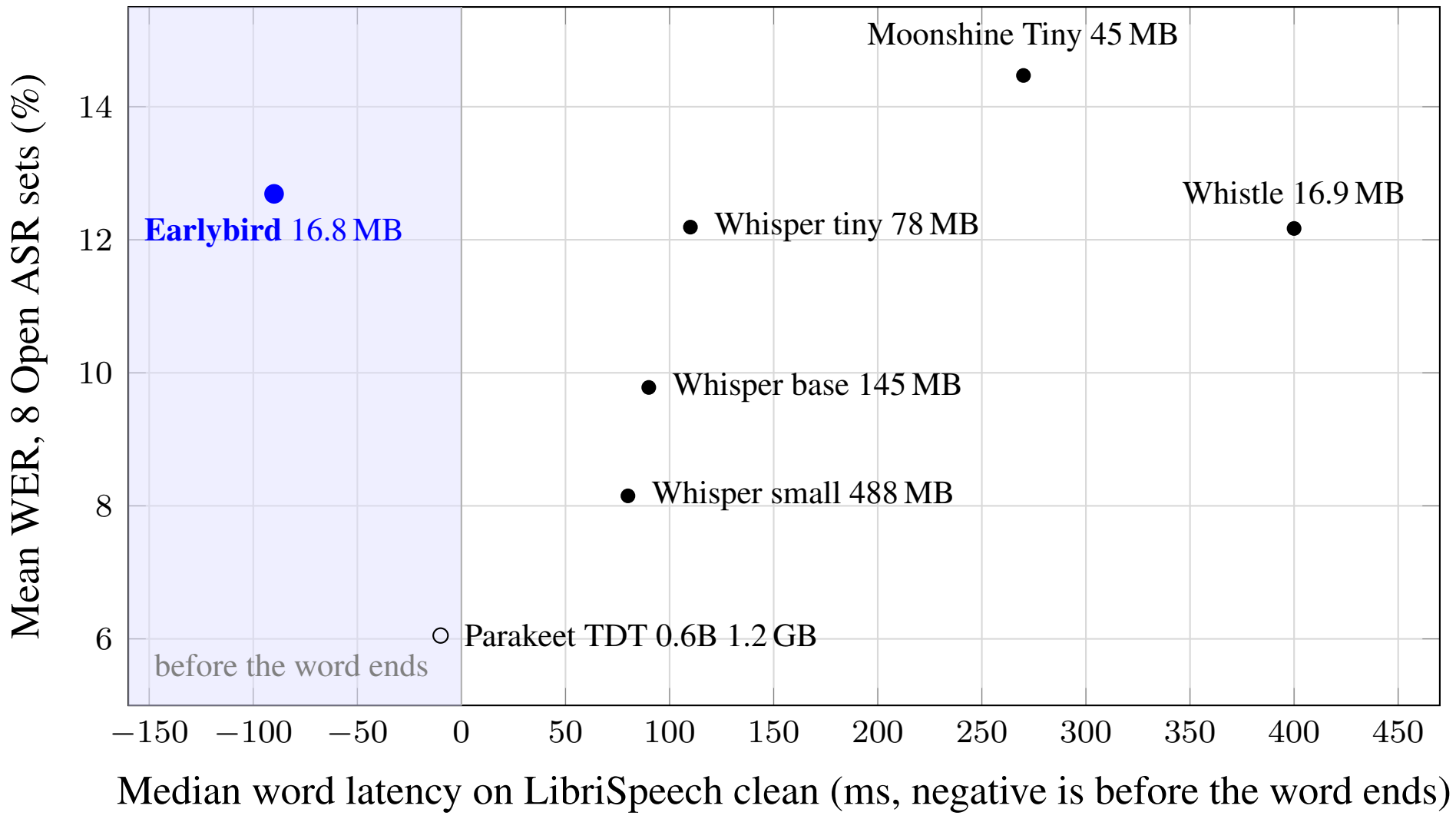
\begin{figure}[t]
\centering
\begin{tikzpicture}
\begin{axis}[width=0.78\linewidth,height=0.46\linewidth,xmin=-160,xmax=470,ymin=5,ymax=15.5,
  xlabel={Median word latency on LibriSpeech clean (ms, negative is before the word ends)},
  ylabel={Mean WER, 8 Open ASR sets (\%)},grid=major,grid style={gray!30},
  tick label style={font=\scriptsize},label style={font=\small}]
\fill[blue!12,fill opacity=0.5] (axis cs:-160,5) rectangle (axis cs:0,15.5);
\draw[gray!60] (axis cs:0,5) -- (axis cs:0,15.5);
\addplot[only marks,mark=*,mark size=2.2pt,blue] coordinates {(-90,12.69)};
\addplot[only marks,mark=*,mark size=1.6pt,black] coordinates {(400,12.17)(270,14.47)(110,12.19)(90,9.78)(80,8.15)};
\addplot[only marks,mark=o,mark size=1.8pt,black] coordinates {(-10,6.05)};
\node[font=\scriptsize,blue,anchor=north,yshift=-2pt] at (axis cs:-90,12.69) {\textbf{\eb{}} 16.8\,MB};
\node[font=\scriptsize,anchor=south,yshift=2pt] at (axis cs:400,12.17) {Whistle 16.9\,MB};
\node[font=\scriptsize,anchor=south,yshift=2pt] at (axis cs:270,14.47) {Moonshine Tiny 45\,MB};
\node[font=\scriptsize,anchor=west,xshift=2pt] at (axis cs:110,12.19) {Whisper tiny 78\,MB};
\node[font=\scriptsize,anchor=west,xshift=2pt] at (axis cs:90,9.78) {Whisper base 145\,MB};
\node[font=\scriptsize,anchor=west,xshift=2pt] at (axis cs:80,8.15) {Whisper small 488\,MB};
\node[font=\scriptsize,anchor=west,xshift=2pt] at (axis cs:-10,6.05) {Parakeet TDT 0.6B 1.2\,GB};
\node[font=\scriptsize,gray,anchor=south west] at (axis cs:-155,5.2) {before the word ends};
\end{axis}
\end{tikzpicture}
\caption{Word latency against accuracy; lower left is better. WER is the mean over the full Open
ASR sets (Whistle: about 200 clips per set), with \eb{} at 4 bits. Latency is on the same 67
LibriSpeech clean utterances for every model except Parakeet, which was re-decoded every 240\ms{} on
150 other utterances (\eb{}: $-80$\ms{} on those). Hollow: published leaderboard WER.}
\label{fig:latacc}
\end{figure}

\begin{figure}[t]
\centering
\begin{tikzpicture}
\begin{axis}[width=0.52\linewidth,height=0.46\linewidth,xmode=log,xmin=4,xmax=2500,ymin=5,ymax=15.5,
  xlabel={Size on disk (MB)},ylabel={Mean WER, 8 Open ASR sets (\%)},grid=major,grid style={gray!30},
  tick label style={font=\scriptsize},label style={font=\small}]
\addplot[only marks,mark=*,mark size=2.2pt,blue] coordinates {(16.8,12.69)(58,11.66)};
\addplot[only marks,mark=*,mark size=1.6pt,black] coordinates {(16.9,12.17)(78,12.19)(145,9.78)(488,8.15)};
\addplot[only marks,mark=o,mark size=1.8pt,black] coordinates {(88,12.01)(280,7.84)(1236,6.05)};
\node[font=\tiny,blue,anchor=south east] at (axis cs:16.8,12.69) {\eb{} 4-bit};
\node[font=\tiny,blue,anchor=north,yshift=-2pt] at (axis cs:58,11.66) {\eb{} fp16};
\node[font=\tiny,anchor=north east] at (axis cs:16.9,12.17) {Whistle};
\node[font=\tiny,anchor=south,yshift=2pt] at (axis cs:78,12.19) {Wh.\ tiny};
\node[font=\tiny,anchor=west,xshift=3pt] at (axis cs:88,12.01) {Moon.\ Tiny};
\node[font=\tiny,anchor=west,xshift=2pt] at (axis cs:145,9.78) {Wh.\ base};
\node[font=\tiny,anchor=west,xshift=2pt] at (axis cs:488,8.15) {Wh.\ small};
\node[font=\tiny,anchor=east,xshift=-2pt] at (axis cs:280,7.84) {Moon.\ Small};
\node[font=\tiny,anchor=south,yshift=2pt] at (axis cs:1236,6.05) {Parakeet};
\end{axis}
\end{tikzpicture}\hfill
\begin{tikzpicture}
\begin{axis}[width=0.52\linewidth,height=0.46\linewidth,xmode=log,xmin=4,xmax=2500,ymin=-160,ymax=470,
  xlabel={Size on disk (MB)},ylabel={Median word latency (ms)},grid=major,grid style={gray!30},
  tick label style={font=\scriptsize},label style={font=\small}]
\fill[blue!12,fill opacity=0.5] (axis cs:4,-160) rectangle (axis cs:2500,0);
\addplot[only marks,mark=*,mark size=2.2pt,blue] coordinates {(16.8,-90)};
\addplot[only marks,mark=*,mark size=1.6pt,black] coordinates {(16.9,400)(45,270)(78,110)(145,90)(488,80)(464,230)(1236,-10)};
\node[font=\tiny,blue,anchor=west,xshift=2pt] at (axis cs:16.8,-90) {\eb{}};
\node[font=\tiny,anchor=west,xshift=2pt] at (axis cs:16.9,400) {Whistle};
\node[font=\tiny,anchor=west,xshift=2pt] at (axis cs:45,270) {Moon.\ Tiny};
\node[font=\tiny,anchor=north,yshift=-2pt] at (axis cs:78,110) {Wh.\ tiny};
\node[font=\tiny,anchor=south,yshift=2pt] at (axis cs:145,90) {Wh.\ base};
\node[font=\tiny,anchor=north,yshift=-2pt] at (axis cs:488,80) {Wh.\ small};
\node[font=\tiny,anchor=south,yshift=2pt] at (axis cs:464,230) {Nemotron};
\node[font=\tiny,anchor=east,xshift=-2pt] at (axis cs:1236,-10) {Parakeet};
\end{axis}
\end{tikzpicture}
\caption{Size against accuracy (left) and against word latency (right); lower left is better.
Full-precision models are shown at their 16-bit size, Nemotron (streaming, 80\ms{} chunks) at its
int8 size. Hollow: published leaderboard WER; filled: our runs. Latency as in
Figure~\ref{fig:latacc}.}
\label{fig:size}
\end{figure}
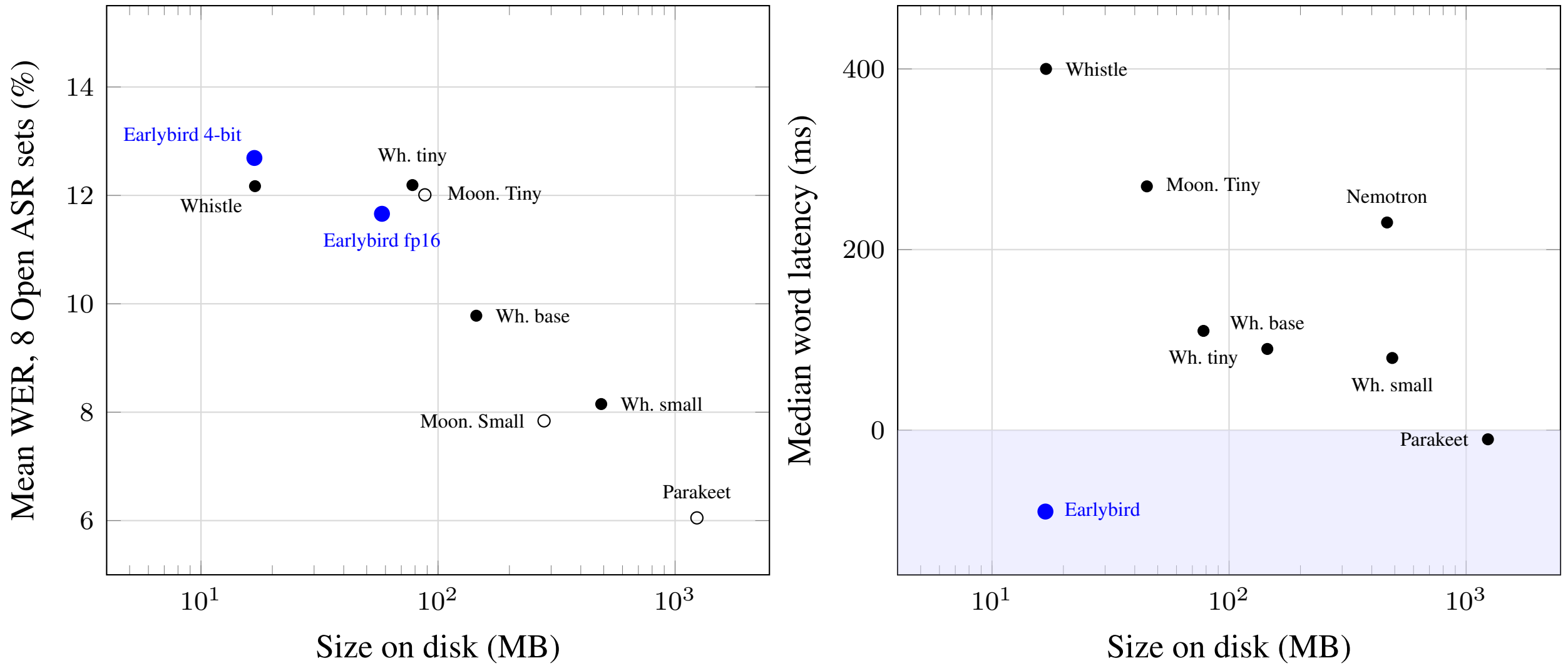

Figures~\ref{fig:latacc} and~\ref{fig:size} place \eb{} among released models of every size, using the per-set results in Sections~\ref{sec:early} and~\ref{sec:openasr}. It is the only
model under 1\,GB whose median word is predicted correctly before it ends. The purpose-built
streaming Nemotron 0.6B~\citep{nemotron} (464\,MB at int8) has a median word latency of +230\ms{} on
the same 67 utterances. Parakeet TDT 0.6B v2~\citep{parakeet}, an offline model 70 times \eb{}'s size
re-decoded every 240\ms{}, reaches $-10$\ms{} with half \eb{}'s error rate. At its own size, \eb{}'s
accuracy is in line with Whistle and Whisper tiny.en and ahead of Moonshine Tiny.

\subsection{Earliness and accuracy}
\label{sec:early}

\begin{table}[t]
\centering
\small
\begin{tabular}{lcccc}
\toprule
Test set & \eb{} & Moonshine Tiny & Whistle & Whisper base \\
\midrule
LibriSpeech clean (67)  & $\mathbf{-90}$ / 2.3 & +270 / 3.9 & +400 / 3.1 & +90 / 2.3 \\
LibriSpeech other (150) & $\mathbf{-60}$ / 9.7 & +300 / 13.2 & +460 / 10.1 & +140 / 8.7 \\
TED-LIUM (150)          & $\mathbf{-68}$ / 5.4 & +298 / 6.8 & +505 / 7.3 & +125 / 4.8 \\
People's Speech (150)   & $\mathbf{-66}$ / 14.5 & +290 / 16.7 & +477 / 14.2 & +127 / 10.5 \\
Earnings-22 (150)       & $\mathbf{-46}$ / 18.4 & +309 / 21.6 & +535 / 17.8 & +142 / 13.3 \\
Synthetic (67)          & $\mathbf{-73}$ / 5.6 & +292 / 7.0 & +424 / 7.5 & +90 / 4.8 \\
\midrule
Revisions per word (LS clean) & 1.29 & 2.07 & 0.16 & 0.39 \\
\bottomrule
\end{tabular}
\caption{Median word latency in ms relative to the end of the word (negative is before) / WER \%,
on the same utterances for every model within a set (count in parentheses). Revisions are words
taken back per spoken word.}
\label{tab:main}
\end{table}

Table~\ref{tab:main} shows that \eb{} predicts the median word correctly before the speaker finishes it
on all six sets, about 360\ms{} sooner than the Moonshine Tiny runtime, 490--580\ms{} sooner than
Whistle and 160--200\ms{} sooner than Whisper base. It makes fewer errors than Moonshine Tiny on
every set. Whistle is ahead on two sets, by at most 0.6 points, and Whisper base is the most accurate of
the four except on LibriSpeech clean, where they tie. The subsets are small, but the medians hold on full test sets: \eb{}'s median word latency is $-90$\ms{} on all 2,620 LibriSpeech clean utterances, $-67$\ms{} on all 1,155 TED-LIUM utterances (Moonshine Tiny runtime: +296\ms{} on the same 1,155), $-63$\ms{} on 500 People's Speech utterances (+301\ms{}) and $-43$\ms{} on 1,000 Earnings-22 utterances (+308\ms{}). The p95 word latency on LibriSpeech clean is
+80\ms{} for \eb{} against +520\ms{} for Moonshine Tiny and +925\ms{} for Whistle. Model size does not
buy earliness for re-decoded Whisper: on the same 67 utterances its median word latency is +110\ms{}
(tiny.en), +90\ms{} (base.en) and +80\ms{} (small.en).

The cost of showing words early is revisions: \eb{} takes back 1.3 words per spoken word on
LibriSpeech clean, fewer than the Moonshine Tiny runtime (2.1) but more than Whistle (0.2) and
Whisper base (0.4), and 2--6 per word on spontaneous or noisy speech.

\begin{table}[t]
\centering
\small
\begin{tabular}{lcccc}
\toprule
Test set & \eb{} & Moonshine Tiny & Whistle & Whisper base \\
\midrule
LibriSpeech clean & \textbf{68.8} & 3.4 & 9.1 & 25.4 \\
TED-LIUM          & \textbf{52.1} & 1.0 & 2.6 & 17.6 \\
People's Speech   & \textbf{31.8} & 1.0 & 2.9 & 11.3 \\
Synthetic         & \textbf{56.9} & 2.2 & 6.8 & 22.0 \\
\bottomrule
\end{tabular}
\caption{Percent of reference words predicted correctly before they end (right, and not changed afterwards),
counting transcript words and guesses output at confidence $\ge 0.5$. Utterances where all four
models produced output: 50, 38, 100 and 50.}
\label{tab:kept}
\end{table}

Median timing hides how many words are actually useful early. Table~\ref{tab:kept} counts words
predicted correctly before they end. Moonshine Tiny and Whistle are
built to wait for each word to end. Whisper base gets some words early because each re-run hears a
little more audio.

\subsection{Standard accuracy: Open ASR Leaderboard}
\label{sec:openasr}

\begin{table}[t]
\centering
\small
\setlength{\tabcolsep}{4pt}
\begin{tabular}{lccccccccc}
\toprule
System & TED & VoxP & LS-c & LS-o & Earn & AMI & Giga & SPGI & Mean \\
\midrule
\eb{} 4-bit (16.8\,MB) & 6.1 & 13.6 & 3.7 & 14.2 & 21.7 & 21.8 & 13.1 & 6.8 & 12.6 \\
\eb{} full precision   & 5.6 & 12.9 & 3.4 & 12.0 & 20.0 & 19.4 & 11.7 & 6.1 & 11.4 \\
Moonshine Tiny runtime & 7.9 & 15.4 & 5.4 & 16.0 & 23.3 & 24.9 & 14.5 & 7.3 & 14.4 \\
Whistle                & 7.7 & 10.8 & 4.1 & 11.3 & 17.9 & 24.0 & 14.6 & 7.1 & 12.2 \\
Whisper tiny.en        & 5.9 & 13.7 & 5.2 & 14.0 & 17.5 & 22.2 & 11.8 & 5.3 & 12.0 \\
Whisper base.en        & 4.9 & 9.8 & 3.8 & 10.2 & 14.4 & 18.5 & 11.6 & 4.0 & 9.7 \\
Whisper small.en       & 4.0 & 8.3 & 2.7 & 7.8 & 11.9 & 16.9 & 8.7 & 3.6 & 8.0 \\
\bottomrule
\end{tabular}
\caption{WER (\%) on the eight English sets of the Open ASR Leaderboard, scored with its
normalizer, every system on the same clips (TED-LIUM in full; about 200 clips of each other set,
because Whistle runs at about real time per CPU core). Final transcripts, not streaming timing.}
\label{tab:openasr}
\end{table}

Table~\ref{tab:openasr} reports final-transcript WER with the Open ASR Leaderboard's
code~\citep{openasr}. Our Whisper base scores match its published entry (TED-LIUM 4.87, identical),
which checks the scoring. \eb{} beats the Moonshine Tiny runtime on all eight sets. The 4-bit
release roughly ties Whistle (12.6 vs.\ 12.2): it wins on five sets and loses on VoxPopuli,
LibriSpeech other and Earnings-22. Quantization costs 1.2 points of mean WER on these sets, more
than on our own test sets. Moonshine Tiny's leaderboard entry (12.0) uses full-precision weights.
The official Moonshine runtime scores 14.4 here. On the full sets, \eb{} averages 11.7 at full precision and 12.7 at 4 bits. Full-precision \eb{} (29M)
is slightly more accurate than Whisper tiny.en (39M, 12.0 here and 12.2 on the full sets), and the 4-bit
release slightly less. Whisper small.en (244M) reaches 8.0.

\subsection{Noise, compute and variance}

\begin{table}[t]
\centering
\small
\begin{tabular}{lccc}
\toprule
Condition & \eb{} & Moonshine Tiny & Whisper base \\
\midrule
White noise, 10\,dB & $\mathbf{-70}$ / 8.7  & +290 / 15.3 & +140 / \textbf{6.5} \\
White noise, 5\,dB  & $\mathbf{-50}$ / 19.5 & +320 / 29.1 & +170 / \textbf{13.4} \\
Chatter, 10\,dB     & $\mathbf{-20}$ / 32.0 & +350 / 41.5 & +150 / \textbf{8.9} \\
Chatter, 5\,dB      & $\mathbf{+130}$ / 65.4 & +490 / 79.0 & +230 / \textbf{29.0} \\
\bottomrule
\end{tabular}
\caption{LibriSpeech test-clean with added noise, 150 utterances: median word latency (ms) / WER \%.}
\label{tab:noise}
\end{table}

\eb{} remains the earliest model at every noise level (Table~\ref{tab:noise}) and is more
accurate than Moonshine Tiny throughout. Its accuracy degrades most under background speech: at
10\,dB, its WER is 32.0\% against 8.9\% for Whisper base.

On one CPU thread (Intel Xeon Platinum 8488C), \eb{} takes 39\ms{} on average per 80\ms{} step
(p95 64\ms{}), about twice as fast as real time, against 47\ms{} for Moonshine Tiny decoded the same way.
Three training runs with different seeds agree within 3\ms{} of median word latency and 0.5 WER
points on every test set.

\begin{table}[t]
\centering
\small
\begin{tabular}{lccc}
\toprule
Variant (LibriSpeech clean, 67 utt.) & Median word latency & Revisions / word & WER \\
\midrule
\eb{} (padding frames)    & $-90$ & 1.29 & 2.3 \\
\eb{}, no padding frames  & $-50$ & 1.10 & 2.0 \\
\eb{} 4-bit               & $-90$ & 1.18 & 1.7 \\
\bottomrule
\end{tabular}
\caption{Effect of padding frames and quantization on the same utterances.}
\label{tab:ablate}
\end{table}

\subsection{Confidence}
\label{sec:conf}

\begin{figure}[t]
\centering
\begin{tikzpicture}
\begin{axis}[width=0.62\linewidth,height=0.5\linewidth,xmin=0,xmax=100,ymin=0,ymax=100,
  xlabel={Stated confidence (\%)},ylabel={Right (\%)},grid=major,grid style={gray!30},
  legend pos=south east,legend style={font=\scriptsize,draw=none,fill=white,fill opacity=0.8,text opacity=1},tick label style={font=\scriptsize},
  label style={font=\small}]
\addplot[gray,dashed,domain=0:100] {x};
\addlegendentry{perfect}
\addplot[blue,mark=*,mark size=1.2pt] coordinates {(6.1,6.7)(15.5,18.2)(25.3,34.6)(35.5,45.4)(45.1,55.8)(55.1,68.6)(64.9,73.2)(75.3,77.9)(85.4,84.9)(96.9,96.0)};
\addlegendentry{LibriSpeech}
\addplot[black,mark=square*,mark size=1.2pt] coordinates {(6.4,3.4)(16.2,18.9)(25.9,31.5)(35.5,41.5)(45.3,52.0)(54.9,62.6)(65.1,70.2)(75.1,73.1)(85.4,83.0)(96.9,94.8)};
\addlegendentry{TED-LIUM}
\addplot[orange!80!black,mark=triangle*,mark size=1.4pt] coordinates {(5.2,2.6)(15.5,16.7)(25.6,25.0)(35.1,38.8)(45.2,47.4)(54.8,57.9)(65.3,62.5)(74.9,66.0)(84.9,74.1)(96.8,90.6)};
\addlegendentry{People's Speech}
\addplot[teal,mark=diamond*,mark size=1.4pt] coordinates {(5.0,5.3)(14.8,10.8)(25.5,29.2)(35.2,39.1)(45.1,52.6)(54.8,63.8)(65.0,64.1)(75.4,72.3)(85.4,82.3)(96.6,94.0)};
\addlegendentry{Synthetic}
\end{axis}
\end{tikzpicture}
\caption{Reliability of the confidence on words still being spoken: every transcript word on
screen whose reference word has not ended, with all words before it right, at every 80\ms{} step,
200 utterances per set.}
\label{fig:rel}
\end{figure}
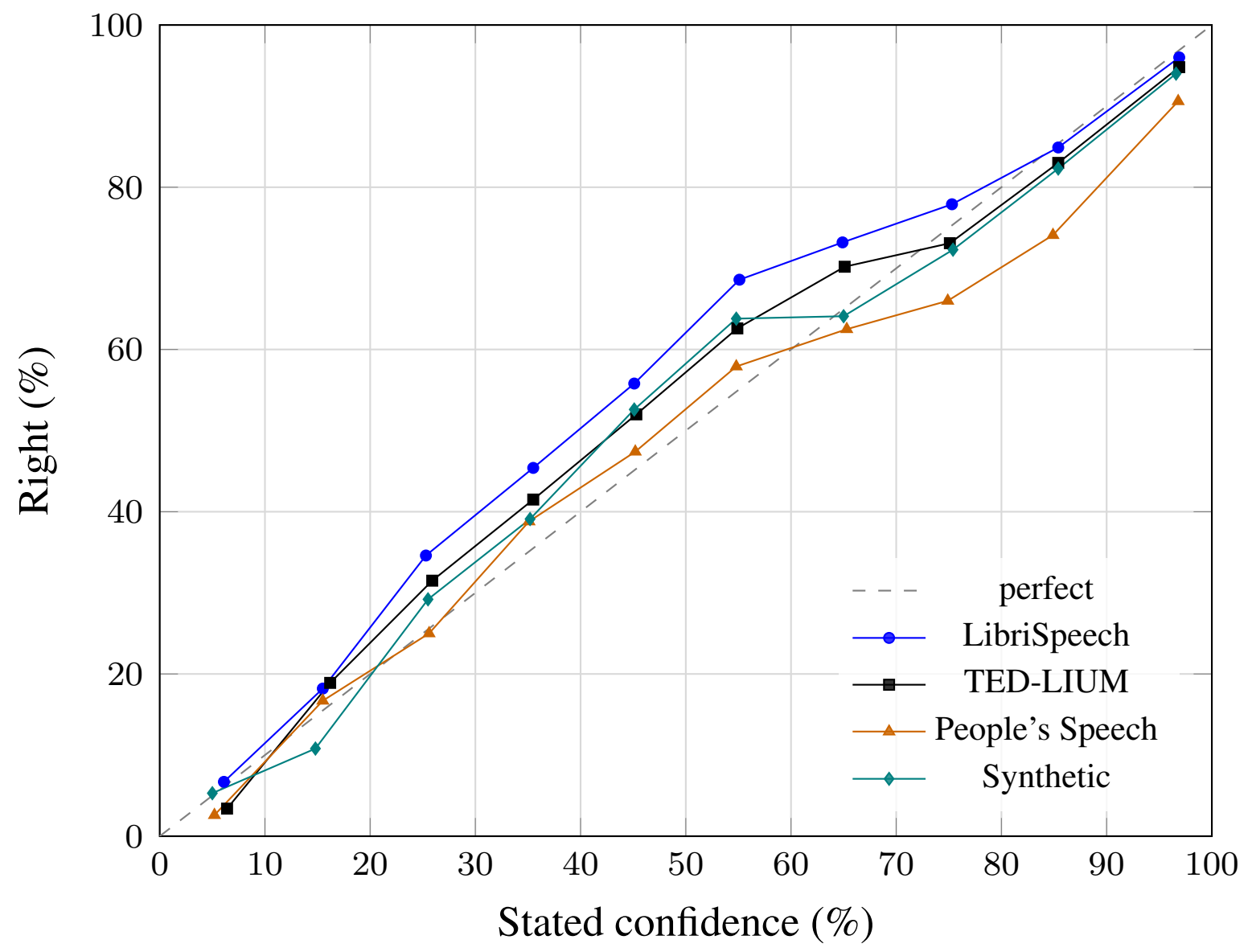

\begin{table}[t]
\centering
\small
\begin{tabular}{lcccc}
\toprule
 & LibriSpeech & TED-LIUM & People's Speech & Synthetic \\
\midrule
Words still being spoken: ECE   & 4.3 & 3.7 & 5.5 & 3.7 \\
\quad right (\%)                & 78.8 & 75.3 & 66.9 & 69.0 \\
Every transcript word: ECE      & 4.3 & 4.7 & 1.4 & 4.0 \\
First guessed word: ECE         & 0.8 & 1.2 & 3.7 & 1.1 \\
Second guessed word: ECE        & 2.0 & 3.4 & 4.4 & 2.1 \\
\bottomrule
\end{tabular}
\caption{Expected calibration error (points, 10 bins) of \eb{}'s raw confidence. No calibration
step is applied.}
\label{tab:ece}
\end{table}

A word output early is only useful if the agent knows when to trust it. Figure~\ref{fig:rel} and
Table~\ref{tab:ece} show that \eb{}'s raw confidence is calibrated for the hardest case, a word the
speaker has not finished: ECE is 3.7--5.5 points, and words output at confidence $\ge 0.9$ are right
91--96\% of the time. Guessed next words are calibrated more tightly (ECE 0.8--3.7). On a set of
spoken commands outside the training domain (SLURP), calibration is worse: ECE 8.8 for words still
being spoken.

\section{Limitations}
\begin{itemize}
  \item \textbf{Accuracy.} Whisper base, 2.5 times larger, is more accurate on every test set
        except LibriSpeech clean, most of all with background chatter.
  \item \textbf{Revisions.} Outputting words early means taking more words back than models that wait,
        especially on spontaneous and noisy speech.
  \item \textbf{Guesses are rare.} At confidence $\ge 0.5$, about one word in seven (12--16\%) is
        guessed right before it ends. Most early words come from the transcript itself.
  \item \textbf{Timing references.} Word end times come from forced aligners, which carry their
        own error of tens of milliseconds. The synthetic set gives exact times and agrees.
  \item \textbf{English only}, and calibration degrades on out-of-domain command speech.
\end{itemize}

\section{Conclusion}
Training a small streaming recognizer on audio cut off at random points, with whole words as
targets, is enough to make it predict words correctly before the speaker finishes them, with a confidence an
application can act on. The resulting model is 16.8\,MB, runs about twice as fast as real time on one CPU
thread, and is about as accurate as other models of its size. On LibriSpeech clean, its median word
is predicted correctly 90\ms{} before it ends, while released streaming models get it 90--400\ms{}
after, and it is the earliest model on all six test sets. 69\% of audiobook words are predicted
correctly before they end, and for words still being spoken its confidence is calibrated, with an
expected calibration error of 3.7 to 5.5. On the eight English Open ASR Leaderboard sets it is more accurate than
Moonshine Tiny on every set and roughly ties Whistle, a model of the same size.
Weights and a loader are released under the MIT license at
\url{https://huggingface.co/sahilmahendrakar/Earlybird-29M}.

\bibliographystyle{unsrtnat}
\bibliography{refs}

\begin{thebibliography}{21}
\providecommand{\natexlab}[1]{#1}
\providecommand{\url}[1]{\texttt{#1}}
\expandafter\ifx\csname urlstyle\endcsname\relax
  \providecommand{\doi}[1]{doi: #1}\else
  \providecommand{\doi}{doi: \begingroup \urlstyle{rm}\Url}\fi

\bibitem[Jeffries et~al.(2024)Jeffries, King, Kudlur, Nicholson, Wang, and
  Warden]{jeffries2024moonshine}
Nat Jeffries, Evan King, Manjunath Kudlur, Guy Nicholson, James Wang, and Pete
  Warden.
\newblock Moonshine: Speech recognition for live transcription and voice
  commands.
\newblock \emph{arXiv preprint arXiv:2410.15608}, 2024.

\bibitem[{Moonshine AI}(2026)]{moonshinestreaming}
{Moonshine AI}.
\newblock Moonshine streaming models and runtime.
\newblock \url{https://github.com/moonshine-ai/moonshine}, 2026.
\newblock MIT license. Accessed November 2026.

\bibitem[Zink et~al.(2024)Zink, Higuchi, Mullov, Waibel, and
  Kobayashi]{zink2025predictive}
Oswald Zink, Yosuke Higuchi, Carlos Mullov, Alexander Waibel, and Tetsunori
  Kobayashi.
\newblock Predictive speech recognition and end-of-utterance detection towards
  spoken dialog systems.
\newblock \emph{arXiv preprint arXiv:2409.19990}, 2024.

\bibitem[Yusuf et~al.(2024)Yusuf, Baskar, Rosenberg, and
  Ramabhadran]{yusuf2024speculative}
Bolaji Yusuf, Murali~Karthick Baskar, Andrew Rosenberg, and Bhuvana
  Ramabhadran.
\newblock Speculative speech recognition by audio-prefixed low-rank adaptation
  of language models.
\newblock In \emph{Proc. Interspeech}, pages 792--796, 2024.

\bibitem[Baumann and Schlangen(2011)]{baumann2011microtiming}
Timo Baumann and David Schlangen.
\newblock Predicting the micro-timing of user input for an incremental spoken
  dialogue system that completes a user's ongoing turn.
\newblock In \emph{Proc. SIGDIAL}, pages 120--129, 2011.

\bibitem[Chang et~al.(2020)Chang, Li, Rybach, He, Li, Sainath, and
  Strohman]{chang2020prefetch}
Shuo-Yiin Chang, Bo~Li, David Rybach, Yanzhang He, Wei Li, Tara~N. Sainath, and
  Trevor Strohman.
\newblock Low latency speech recognition using end-to-end prefetching.
\newblock In \emph{Proc. Interspeech}, pages 1962--1966, 2020.

\bibitem[Schwarz et~al.(2023)Schwarz, He, Van~Segbroeck, Hethnawi, and
  Rastrow]{schwarz2023personalized}
Andreas Schwarz, Di~He, Maarten Van~Segbroeck, Mohammed Hethnawi, and Ariya
  Rastrow.
\newblock Personalized predictive {ASR} for latency reduction in voice
  assistants.
\newblock In \emph{Proc. Interspeech}, 2023.

\bibitem[Yu et~al.(2021)Yu, Chiu, Li, Chang, Sainath, He, Narayanan, Han,
  Gulati, Wu, and Pang]{yu2021fastemit}
Jiahui Yu, Chung-Cheng Chiu, Bo~Li, Shuo-yiin Chang, Tara~N. Sainath, Yanzhang
  He, Arun Narayanan, Wei Han, Anmol Gulati, Yonghui Wu, and Ruoming Pang.
\newblock {FastEmit}: Low-latency streaming {ASR} with sequence-level emission
  regularization.
\newblock In \emph{Proc. ICASSP}, 2021.

\bibitem[Mach{\'a}{\v{c}}ek et~al.(2023)Mach{\'a}{\v{c}}ek, Dabre, and
  Bojar]{machacek2023whisperstreaming}
Dominik Mach{\'a}{\v{c}}ek, Raj Dabre, and Ond{\v{r}}ej Bojar.
\newblock Turning {Whisper} into real-time transcription system.
\newblock In \emph{Proc. IJCNLP-AACL: System Demonstrations}, 2023.

\bibitem[Panayotov et~al.(2015)Panayotov, Chen, Povey, and
  Khudanpur]{panayotov2015librispeech}
Vassil Panayotov, Guoguo Chen, Daniel Povey, and Sanjeev Khudanpur.
\newblock Librispeech: An {ASR} corpus based on public domain audio books.
\newblock In \emph{Proc. ICASSP}, 2015.

\bibitem[McAuliffe et~al.(2017)McAuliffe, Socolof, Mihuc, Wagner, and
  Sonderegger]{mcauliffe2017mfa}
Michael McAuliffe, Michaela Socolof, Sarah Mihuc, Michael Wagner, and Morgan
  Sonderegger.
\newblock Montreal forced aligner: Trainable text-speech alignment using
  {Kaldi}.
\newblock In \emph{Proc. Interspeech}, 2017.

\bibitem[Galvez et~al.(2021)Galvez, Diamos, Ciro, Cer{\'o}n, Achorn, Gopi,
  Kanter, Lam, Mazumder, and Reddi]{galvez2021peoples}
Daniel Galvez, Greg Diamos, Juan Ciro, Juan~Felipe Cer{\'o}n, Keith Achorn,
  Anjali Gopi, David Kanter, Maximilian Lam, Mark Mazumder, and Vijay~Janapa
  Reddi.
\newblock The people's speech: A large-scale diverse {English} speech
  recognition dataset for commercial usage.
\newblock In \emph{NeurIPS Datasets and Benchmarks}, 2021.

\bibitem[{hexgrad}(2025)]{kokoro}
{hexgrad}.
\newblock Kokoro-82{M}.
\newblock \url{https://huggingface.co/hexgrad/Kokoro-82M}, 2025.

\bibitem[Hernandez et~al.(2018)Hernandez, Nguyen, Ghannay, Tomashenko, and
  Est{\`e}ve]{hernandez2018tedlium}
Fran{\c{c}}ois Hernandez, Vincent Nguyen, Sahar Ghannay, Natalia Tomashenko,
  and Yannick Est{\`e}ve.
\newblock {TED-LIUM} 3: Twice as much data and corpus repartition for
  experiments on speaker adaptation.
\newblock In \emph{Proc. SPECOM}, 2018.

\bibitem[Del~Rio et~al.(2022)Del~Rio, Ha, McNamara, Miller, and
  Chandra]{delrio2022earnings}
Miguel Del~Rio, Peter Ha, Quinten McNamara, Corey Miller, and Shipra Chandra.
\newblock Earnings-22: A practical benchmark for accents in the wild.
\newblock \emph{arXiv preprint arXiv:2203.15591}, 2022.

\bibitem[Guo et~al.(2017)Guo, Pleiss, Sun, and Weinberger]{guo2017calibration}
Chuan Guo, Geoff Pleiss, Yu~Sun, and Kilian~Q. Weinberger.
\newblock On calibration of modern neural networks.
\newblock In \emph{Proc. ICML}, 2017.

\bibitem[{Cactus Compute}(2026)]{whistle}
{Cactus Compute}.
\newblock Whistle.
\newblock \url{https://huggingface.co/Cactus-Compute/whistle}, 2026.
\newblock Engine build of 2026-10-19.

\bibitem[Radford et~al.(2023)Radford, Kim, Xu, Brockman, McLeavey, and
  Sutskever]{radford2023whisper}
Alec Radford, Jong~Wook Kim, Tao Xu, Greg Brockman, Christine McLeavey, and
  Ilya Sutskever.
\newblock Robust speech recognition via large-scale weak supervision.
\newblock In \emph{Proc. ICML}, 2023.

\bibitem[{NVIDIA}(2026)]{nemotron}
{NVIDIA}.
\newblock Nemotron speech streaming {EN} 0.6b.
\newblock \url{https://github.com/k2-fsa/sherpa-onnx}, 2026.
\newblock Run as the int8 sherpa-onnx streaming export. Accessed November 2026.

\bibitem[{NVIDIA}(2025)]{parakeet}
{NVIDIA}.
\newblock Parakeet {TDT} 0.6b v2.
\newblock \url{https://huggingface.co/nvidia/parakeet-tdt-0.6b-v2}, 2025.
\newblock Run as the int8 sherpa-onnx export. Accessed November 2026.

\bibitem[{Hugging Face}(2026)]{openasr}
{Hugging Face}.
\newblock Open {ASR} leaderboard.
\newblock \url{https://huggingface.co/spaces/hf-audio/open_asr_leaderboard},
  2026.
\newblock Evaluation code:
  \url{https://github.com/huggingface/open_asr_leaderboard}.

\end{thebibliography}
\end{document}